\documentclass[twocolumn,english]{paper}
\usepackage{ae,aecompl}
\usepackage{helvet}

\usepackage[T1]{fontenc}
\usepackage[latin9]{inputenc}
\usepackage{color}
\usepackage{amssymb}
\usepackage{graphicx}
\usepackage{esint}
\usepackage{babel}

\begin{document}

\title{Non-ergodic metallic and insulating phases of Josephson junction chains}
\author{M. Pino\textsuperscript{a}, B.L. Altshuler\textsuperscript{b} and L.B. Ioffe\textsuperscript{a,c}\\
\textsuperscript{a} \normalsize\itshape{Department of Physics and Astronomy, Rutgers The State University of New Jersey,}\\
\normalsize\itshape{136 Frelinghuysen rd, Piscataway, 08854 New Jersey, USA}\\
\textsuperscript{b} \normalsize\itshape{Physics Department, Columbia University, New York 10027, USA}\\
\textsuperscript{c} \normalsize\itshape{ LPTHE, CNRS UMR 7589, Boite 126, 4 Jussieu, 75252 Paris Cedex 05, France}\\}
\maketitle

\begin{abstract}
\textbf{Strictly speaking the laws of the conventional Statistical
Physics, based on the Equipartition Postulate\cite{Gibbs1902} and
Ergodicity Hypothesis\cite{Boltzmann1964}, apply only in the presence
of a heat bath. Until recently this restriction was not important
for real physical systems: a weak coupling with the bath was believed
to be sufficient. However, the progress in both quantum gases and
solid state coherent quantum devices demonstrates that the coupling
to the bath can be reduced dramatically. To describe such systems
properly one should revisit the very foundations of the Statistical
Mechanics. We examine this general problem for the case of the Josephson
junction chain and show that it displays a novel high temperature
non-ergodic phase with finite resistance. With further increase of
the temperature the system undergoes a transition to the fully localized
state characterized by infinite resistance and exponentially long
relaxation. }
\end{abstract}
The remarkable feature of the closed quantum systems is the appearance
of Many-Body Localization (MBL)\cite{Basko2006}: under certain conditions
the states of a many-body system are localized in the Hilbert space
resembling the celebrated Anderson Localization \cite{Anderson1958}
of single particle states in a random potential. MBL implies that
macroscopic states of an isolated system depend on the initial conditions
i.e. the time averaging does not result in equipartition distribution
and the entropy never reaches its thermodynamic value. Variation of
macroscopic parameters, e.g. temperature, can delocalize the many
body state. However, the delocalization does not imply the recovery
of the equipartition. Such a non-ergodic behavior in isolated physical
systems is the subject of this Letter. 

We argue that regular Josephson junction arrays (JJA) under the conditions
that are feasible to implement and control experimentally demonstrate
both MBL and non-ergodic behavior. A great advantage of the Josephson
circuits is the possibility to disentangle them from the environment
as was demonstrated by the quantum information devices.\cite{Devoret2013}
At low temperatures the conductivity $\sigma$ of JJA is finite (below
we call such behavior metallic), while as $T\rightarrow0$ JJA becomes
either a superconductor ($\sigma\rightarrow\infty)$ or an insulator
($\sigma\rightarrow0)$\cite{Efetov1980,Fazio2001}. We predict that
at a critical temperature $T=T_{c}$ JJA undergoes a true phase transition
into a MBL insulator ($\sigma=0$ for $T>T_{c}$). Remarkably already
in the metallic state JJA becomes nonergodic and can not be properly
described by the conventional Statistical Mechanics. 

JJA is characterized by the set of phases $\{\phi_{i}\}$ and charges$\{q_{i}\}$
of the superconducting islands, $\phi_{i}$ and $q_{i}$ for each
$i$ are canonically conjugated. The Hamiltonian $H$ is the combination
of the charging energies of the islands with the Josephson coupling
energies. Assuming that the ground capacitance of the islands dominates
their mutual capacitances (this assumption is not crucial for the
qualitative conclusions) we can write $H$\textcolor{green}{{} }as 
\begin{equation}
H=\sum_{i}\left[\frac{1}{2}E_{C}q_{i}^{2}+E_{J}(1-\cos\phi_{i})\right]\label{eq:H}
\end{equation}

The ground state of the model (\ref{eq:H}) is determined by the ratio
of the Josephson and charging energies, $E_{J}/E_{C}$ that controls
the strength of quantum fluctuations: JJA is an insulator at $E_{J}/E_{C}<\eta$
and a superconductor at $E_{J}/E_{C}>\eta$ \cite{Efetov1980,Fazio2001}
with $\eta\approx0.63$ (see Supplemental materials). The quantum
transition at $E_{J}/E_{C}=\eta$ belongs to the Berezinsky-Kosterlitz-Thouless
universality class \cite{Kosterlitz2013}. Away from the ground state
in addition to $E_{J}/E_{C}$ there appears dimensionless parameter
$U/E_{J}$ where $U$ is the energy per superconducting island ($U=T$
in the thermodynamic equilibrium at $T\ll E_{J}$).

\begin{figure}[t!]
\includegraphics[width=0.9\columnwidth]{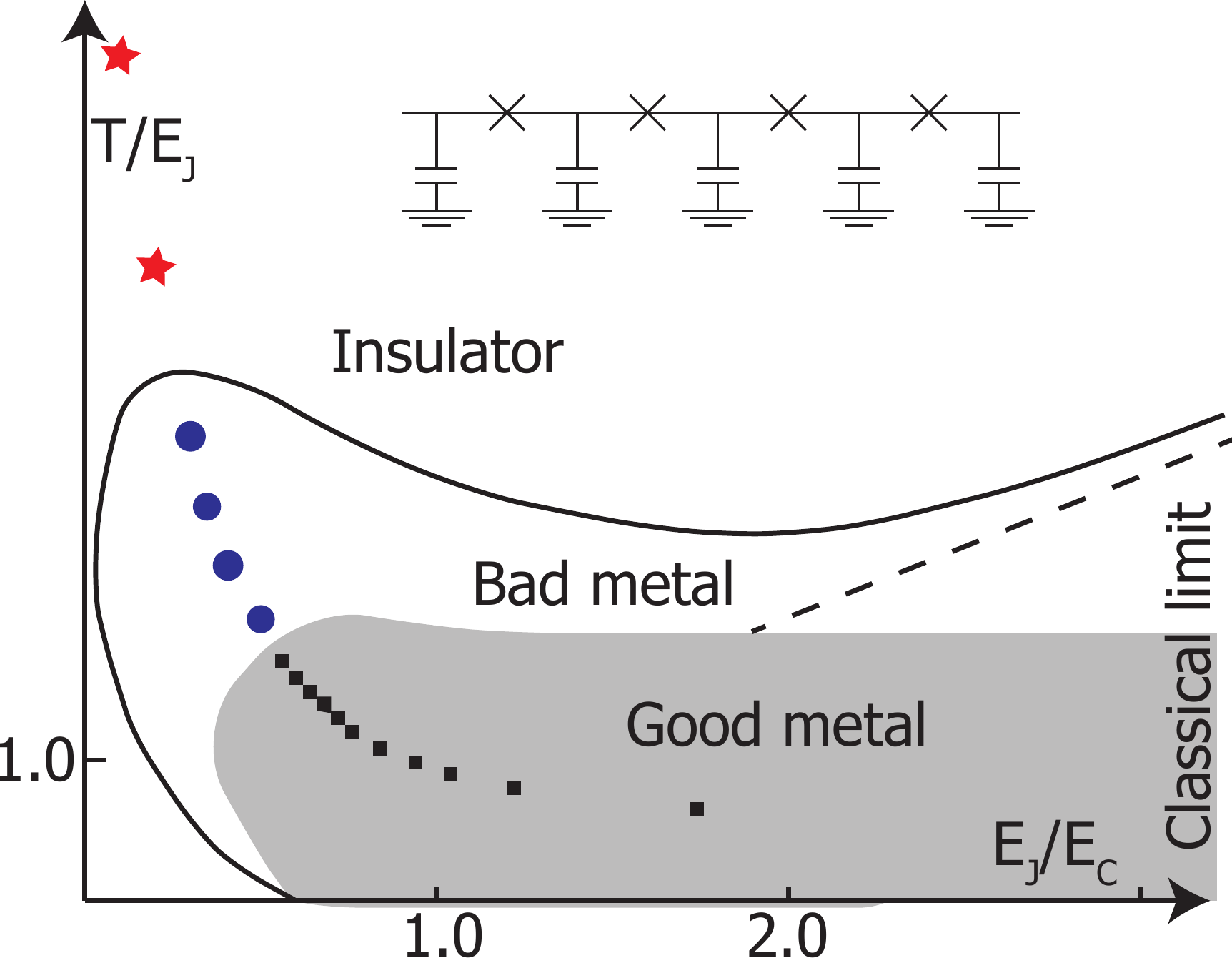}

\protect\caption{Phase diagram of one dimensional Josephson junction array. The MBL
phase transition separates the non-ergodic bad metal with exponentially
large but finite resistance from the insulator with infinite resistance.
Cooling the non-ergodic bad metal transforms it into a good ergodic
metal. The points show approximate positions of the effective $T/E_{J}$
for the quantum problem with a finite number of charging states. The
red stars indicate insulator, blue circles bad metal, and squares
good metal. \label{fig:Phase-diagram-of}}
\end{figure}

The main qualitative finding of this paper is the appearance of a
non-ergodic and highly resistive ``bad metal'' phase at high temperatures,
$T/E_{J}>1$, which at $T\geq T_{c}\approx E_{J}^{2}/E_{C}$ undergoes
the transition to the MBL insulator. In contrast to the $T=0$ behavior,
these results are robust, e.g. are insensitive to the presence of
static random charges. The full phase diagram in the variables $E_{J}/E_{C}$
, $T/E_{J}$ is shown in Fig. \ref{fig:Phase-diagram-of}. We confirmed
numerically that the bad metal persists in the classical limit although
$T_{c}\rightarrow\infty$; it is characterized by the exponential
growth of the resistance with $T$ and violation of thermodynamic
identities. We support these findings by semi-quantitative theoretical
arguments. Finally, we present the results of numerical diagonalization
and tDMRG (time Density Matrix Renormalization Group\cite{White92})
of \emph{quantum} systems that demonstrate both the non-ergodic bad
metal and the MBL insulator. 

It is natural to compare the non-ergodic state of JJA with a conventional
glass characterized by infinitely many metastable states. The glass
entropy does not vanish at $T=0$, i. e. when heated from $T=0$ to
the melting temperature the glass releases less entropy than the crystal
(Kauzmann paradox \cite{Kauzmann1948}). Similarly to glasses JJA
demonstrates non-ergodic behavior in both quantum and classical regimes.
However the ergodicity violation emerges as high rather than low temperatures
transforming Kauzmann paradox into an apparent temperature divergence
(see below). 

\emph{Qualitative arguments for MBL transition.} In a highly excited
state $U\gg E_{J}$ the charging energy dominates: $E_{C}q^{2}\sim U\gg E_{J}$.
Accordingly, the value of the charge, $|q_{i}$|, and charge difference
on neighboring sites, $\delta q_{i}=|q_{i}-q_{i+1}|$ are of the order
of $q\sim\delta q\sim\sqrt{U/E_{C}}$. The energy cost of a unit charge
transfer between two sites $\delta E\sim\sqrt{UE_{c}}$ exceeds the
matrix element of the charge transfer, $E_{J}/2$, as long as $U\gg U_{MBL}=E_{J}^{2}/E_{C}$.
According to \cite{Anderson1958,Basko2006} the system is a non-ergodic
MBL insulator under this condition. Thus, we expect the transition
to MBL phase at $T_{c}/E_{J}\propto E_{J}/E_{C}$ with the numerical
prefactor close to unity (see supplemental materials). 

If $E_{J}/E_{C}\gg1$ and $U\sim T\ll E_{J}$ the conductivity limited
by thermally activated phase slips is exponentially large, $\sigma\sim\exp(E_{J}/T)$\cite{Tinkham1996,Schmidt2002}.
As we show below, at $T\gg E_{J}$ in the metallic phase conductivity
is exponentially small, $\sigma\sim\exp(-T/E_{J})$, even far from
the transition, $E_{J}\ll T\ll E_{J}^{2}/E_{C}$. The resistance in
this state can exceed $R_{Q}=h/(2e)^{2}$dramatically and still display
the ``metallic'' temperature behavior ($dR/dT>0$). 

\emph{Classical regime} is realized at $E_{C}\rightarrow0$ for fixed
$E_{J}$ and $T$. One can express the charge of an island $q$ through
the dimensionless time $\tau=\sqrt{E_{J}E_{C}}t$ as $q=\sqrt{E_{J}/E_{C}}d\phi/d\tau$.
Since $q\sim\sqrt{E_{J}/E_{C}}\gg1$. one can neglect the charge quantization
and use the equations of motion
\begin{equation}
\frac{d^{2}\phi_{i}}{d\tau^{2}}=\sin(\phi_{i+1}-\phi_{i})+\sin(\phi_{i-1}-\phi_{i}),\label{eq:Eq_of_motion}
\end{equation}
Here $i=1,...,L$ and the boundary conditions are $\phi_{0}=\phi_{L+1}=0$.
We solve these equations for the various initial conditions corresponding
to a given total energy and compute the energy $U_{S}$ contained
in a part of the whole chain of the length $1\ll l\ll L$ as a function
of $\tau$. 

The ergodicity implies familiar thermodynamic identities. e.g. 
\begin{equation}
\left(\left\langle U_{S}^{2}\right\rangle -\left\langle U_{S}\right\rangle ^{2}\right)/T^{2}=d\left\langle U_{S}\right\rangle /dT\label{eq:U^2}
\end{equation}
This relation between the average energy of the subsystem, $\left\langle U_{S}\right\rangle $
and its second moment $\left\langle U_{S}^{2}\right\rangle $ turns
out to be invalid for a bad metal. To demonstrate this we evaluated
the average energy per site in this subsystem, $u=\overline{\left\langle U_{S}\right\rangle _{\tau}}/(E_{J}l)$
and the temporal fluctuations of this energy, $w_{\tau}=\overline{\left(\left\langle U_{S}^{2}\right\rangle _{\tau}-\left\langle U_{S}\right\rangle _{\tau}^{2}\right)}/(E_{J}^{2}l)$.
Here $\left\langle \ldots\right\rangle _{\tau}$ and the bar denote
correspondingly averaging over the time %
\footnote{Given the evolution time $\tau_{av}$ $\left\langle \ldots\right\rangle _{\tau}$
averaging means averaging over the time interval $\tau_{ev}$ after
initial evolution for time $\tau_{ev}$.%
} and over the ensemble of the initial conditions. 

From (\ref{eq:H}) it follows that $u=T/(2E_{J})$ at $T\gg E_{J}$
($u\left(T\right)$-function is evaluated for arbitrary $T/E_{J}$
in supplemental materials). One can thus rewrite (\ref{eq:U^2}) as
\begin{equation}
w=\frac{T^{2}}{E_{J}}\frac{du}{dT}\approx2u^{2}\label{eq:w}
\end{equation}

\begin{figure}[t!]
\includegraphics[width=0.9\columnwidth]{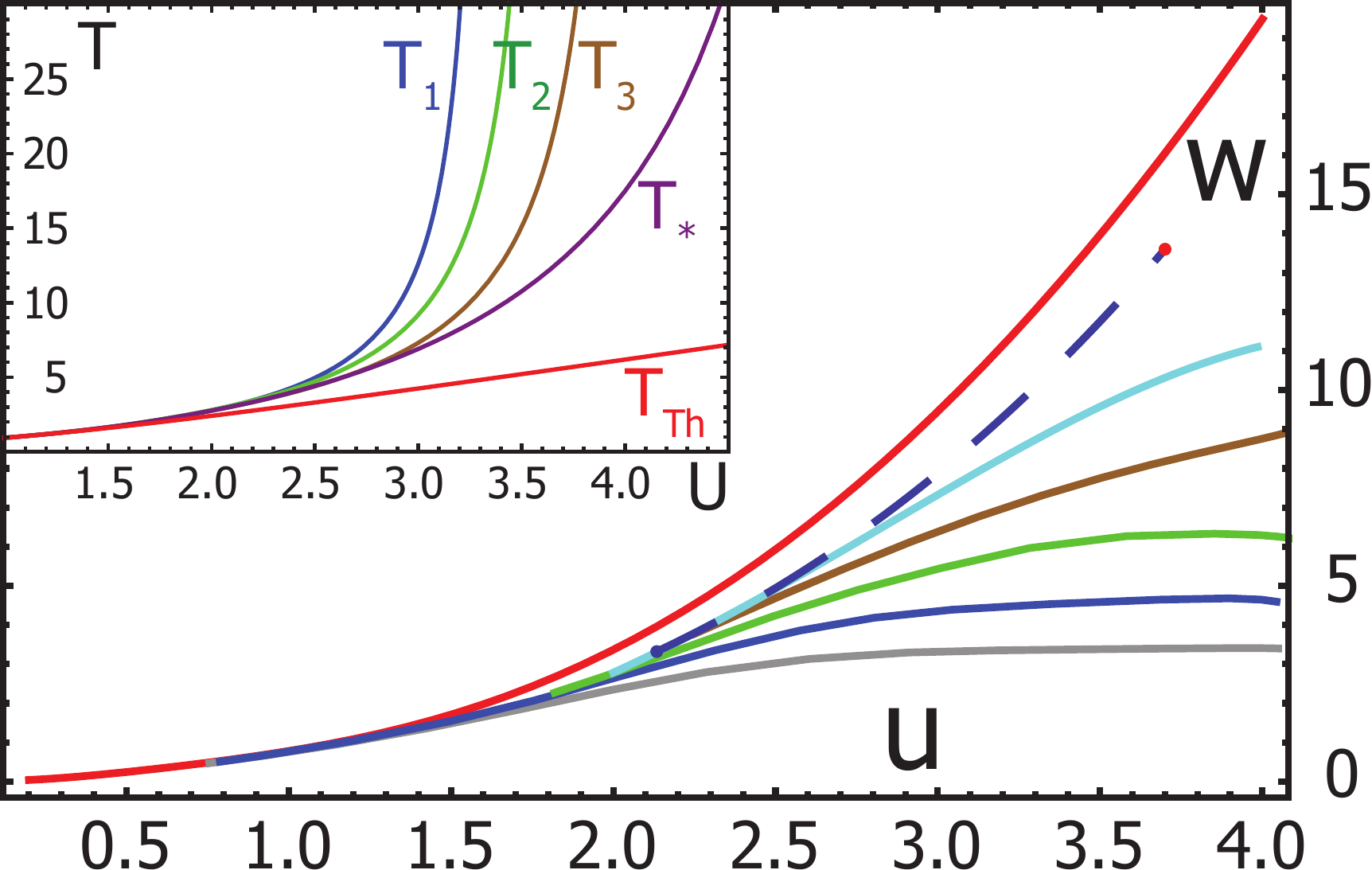}

\protect\caption{Energy fluctuations as a function of average energy per island for
different evolution times ($\tau_{ev}=1,\,2,\,4,\,8,16\,\times10^{4}$
) in a small subsystem; the dashed line is the extrapolation to infinite
times as explained in the text. A single point (red) obtained by direct
computations up to $\tau_{ev}=2\,10^{6}$ at which time scales the
time dependence practically disappears for $L\lesssim100$. The upper
(red) curve corresponds to the thermodynamic identity (\ref{eq:U^2}).
The insert shows the effective temperature defined by the energy fluctuations
determined at different time scales: $\tau_{ev}=\,2,\,4,\,8\times10^{4}$
($T_{1}-T_{3}$ respectively) and by their extrapolation to infinite
times ($T_{*}$), their comparison with the temperature expected in
thermodynamic equilibrium, $T_{Th}$. \label{fig:Energy-fluctuations-as} }
\end{figure}

Results of the numerical solution of (\ref{eq:Eq_of_motion}) are
compared with (\ref{eq:U^2}) in Fig. \ref{fig:Energy-fluctuations-as}.
For any given evolution time, $\tau_{ev}$, the computed $w_{\tau}\left(u\right)$-
dependence saturates instead of increasing as $u^{2}$ (\ref{eq:w}).
At a fixed $u$, $w_{\tau}\left(u\right)$ increases with time extremely
slowly. Below we argue that $w_{\tau}\left(u\right)$ does not reach
its thermodynamic value even at $\tau\to\infty$. Violation of the
thermodynamic identity implies that temperature is ill defined, so
the average energy $u$ rather than $T$ is the proper control parameter.
The effective temperature defined as $T_{*}(u)=E_{J}/\int_{0}^{u}du/w(u)$,
is shown in the insert to Fig. \ref{fig:Energy-fluctuations-as}. 

\emph{Qualitative interpretation. }Large dispersion of charges on
adjacent islands, $i,i+1$ at $u\gg1$ implies quick change of the
phase differences, $\phi_{i}-\phi_{i+1}$, with time. Typical current
between the two neighbors $E_{J}\sin(\phi_{i}-\phi_{i+1})$ time-averages
almost to zero. However, accidentally the frequencies, $\omega_{i}=d\phi_{i}/dt$,
get close. In the classical limit the difference $\omega_{i}-\omega_{i+1}$
can be arbitrary small. Such pair of islands is characterized by one
periodic in time phase difference. Contrarily, three consecutive islands
with close frequencies $|\omega_{i}-\omega_{i+1}|\sim|\omega_{i}-\omega_{i-1}|\lesssim1$
experience chaotic dynamics that contains arbitrary small frequencies,
similarly to work \cite{Chirikov1979}. For uncorrelated frequencies
with variances $u\gg1$, a triad of islands $(i-1,\, i,\, i+1)$ is
chaotic with the probability $1/u\ll1$, i.e. such triads are separated
by large quiet regions of a typical size $r_{t}\sim u\gg1$. The low
frequency noise generated by a chaotic triad decreases exponentially
deep inside a quiet region. Provided that $\omega\ll\omega_{i+m}$
, $m=0\ldots r$ the superconducting order parameter $z_{i}(\omega)=\int d\tau\exp(i\phi_{i}+i\omega\tau)$
at site $i$ satisfies the recursion relation
\begin{equation}
z_{i+r}(\omega)=z_{i}(\omega)\prod_{m=0}^{r-1}\frac{1}{2\omega_{i+m}^{2}}\label{eq:z_i+k}
\end{equation}
which implies the log-normal distribution for the resistances $R_{j,j+r}$
of quiet regions (see Supplemental materials):
\begin{eqnarray}
\left\langle \ln^{2}\left(R/R_{t}\right)\right\rangle  & = & \ln R_{t}\label{eq:Ln^2R}\\
\ln R_{t}=\left\langle \ln R\right\rangle  & = & u\ln\left(u\right)\label{eq:LnR}
\end{eqnarray}
where $R_{t}$ is the typical resistance of a quiet region. The resistance
of the whole array is the sum of the resistances of the quiet regions.
The mean number of these regions in the chain equals to $N=L/r_{t}\gg1$,
its fluctuations being negligible. For the log normal distribution
the average resistance of a quiet region, $\left\langle R\right\rangle $,
is given by$\left\langle R\right\rangle =R_{t}^{3/2}$. For the resistivity,
$\rho$, we thus have $\rho=N\left\langle R\right\rangle /L=R_{t}^{3/2}/r_{t}$.
According to (\ref{eq:LnR}) 
\begin{equation}
\ln\rho r_{t}\approx\frac{3}{2}\left(\gamma+\ln u\right)r_{t}\approx u(\ln u+\gamma)\label{eq:Lnrhor}
\end{equation}
where $\gamma=0.577$ is Euler constant.

\begin{figure}[t!]
\includegraphics[width=0.9\columnwidth]{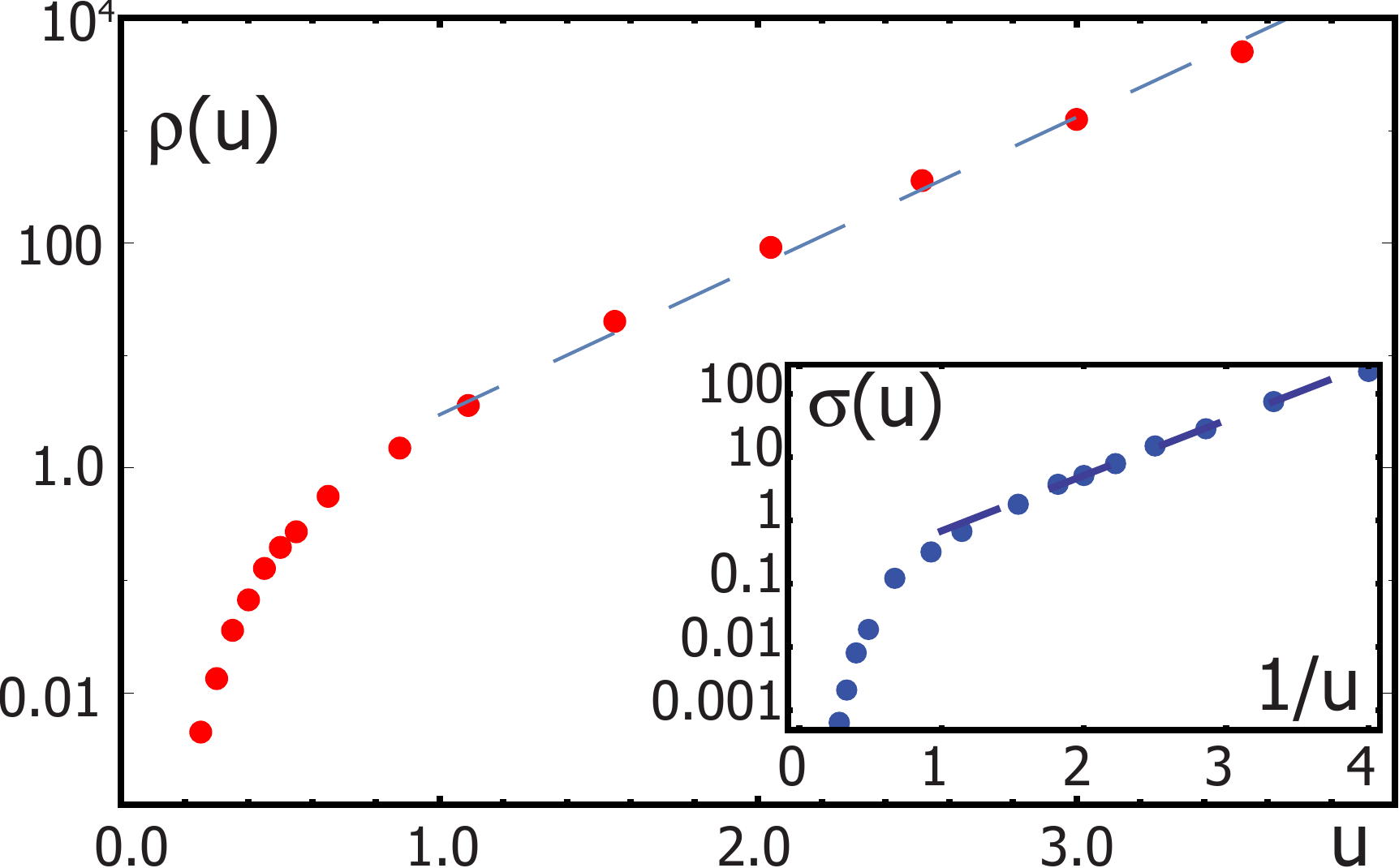}

\protect\caption{Resistivity as a function of the internal energy. At high energies
resistivity is exponentially large due to large regions of almost
frozen charges (see text). Insert: at low energies (temperatures)
exponentially large conductivity is limited by exponentially rare
phase slips. High $u$ points require computation times $\tau_{ev}\sim10^{8}.$
\label{fig:Resistance-as-a}}
\end{figure}

In order to determine the current caused by voltage $V$ across the
chain we solved the equations (\ref{eq:Eq_of_motion}) with modified
boundary conditions, $\phi_{0}=0$ , $\phi_{L+1}=Vt$. The results
confirm the prediction (\ref{eq:Lnrhor}), see Fig. \ref{fig:Resistance-as-a}.
The range of the resistances set by realistic computation time is
too small to detect the logarithmic factor in (\ref{eq:Lnrhor},\ref{eq:LnR}),
however, a relatively large slope, $d\ln\rho/du\approx3.0$ at $u=3.5$
is consistent with (\ref{eq:Lnrhor}) that gives $dln\rho/du\approx2.5+1.5\ln u-1/u$. 

The qualitative picture of triads separated by log-normally distributed
resistances of silent regions allows one to understand the long time
relaxation of $w_{\tau}(u)$ in the subsystem of length $l$ (see
Fig. \ref{fig:Energy-fluctuations-as}). Each resistance can be viewed
as a barrier with a tunneling rate $\sim1/R$. For a given time $\tau$
the barriers with $\tau\ll R$ can be considered impenetrable, whilst
the barriers with $\tau\gg R$ can be neglected. As a result, the
barriers with $R\geq\tau$ break the system into essentially independent
quasiequilibrium regions (QER) of the typical size 
\begin{equation}
l_{\tau}\sim\frac{r_{t}}{\sqrt{2\pi\ln R_{t}}}\exp\left[\frac{\ln^{2}(\tau/R_{t})}{2\ln R_{t}}\right].\label{eq:l_tau}
\end{equation}
If $l\gg r_{t}$ and $\tau\lesssim\exp\left[\sqrt{\ln(l/r_{t})\ln R_{\tau}}\right]$
the subsystem contains $l_{\tau}/l\gg1$ QER, so that $w\propto l_{\tau}/l$.
At longer times, $l_{\tau}\gg l$, the subsystem is in equilibrium
with a particular QER and $w\propto l/l_{\tau}$. The full dependence
on time can be interpolated as 
\begin{equation}
w_{\tau}=\frac{w_{\infty}}{1+\beta\exp\left(-\alpha\ln^{2}\left(\tau/\tau_{0}\right)\right)}\label{eq:w_tau}
\end{equation}
where $\alpha=(2\ln R_{t})^{-1}$, $\beta=\sqrt{\pi/\alpha}(l/r_{t})$
and $\tau_{0}=R_{t}$. 

\begin{figure}[t!]
\includegraphics[width=0.9\columnwidth]{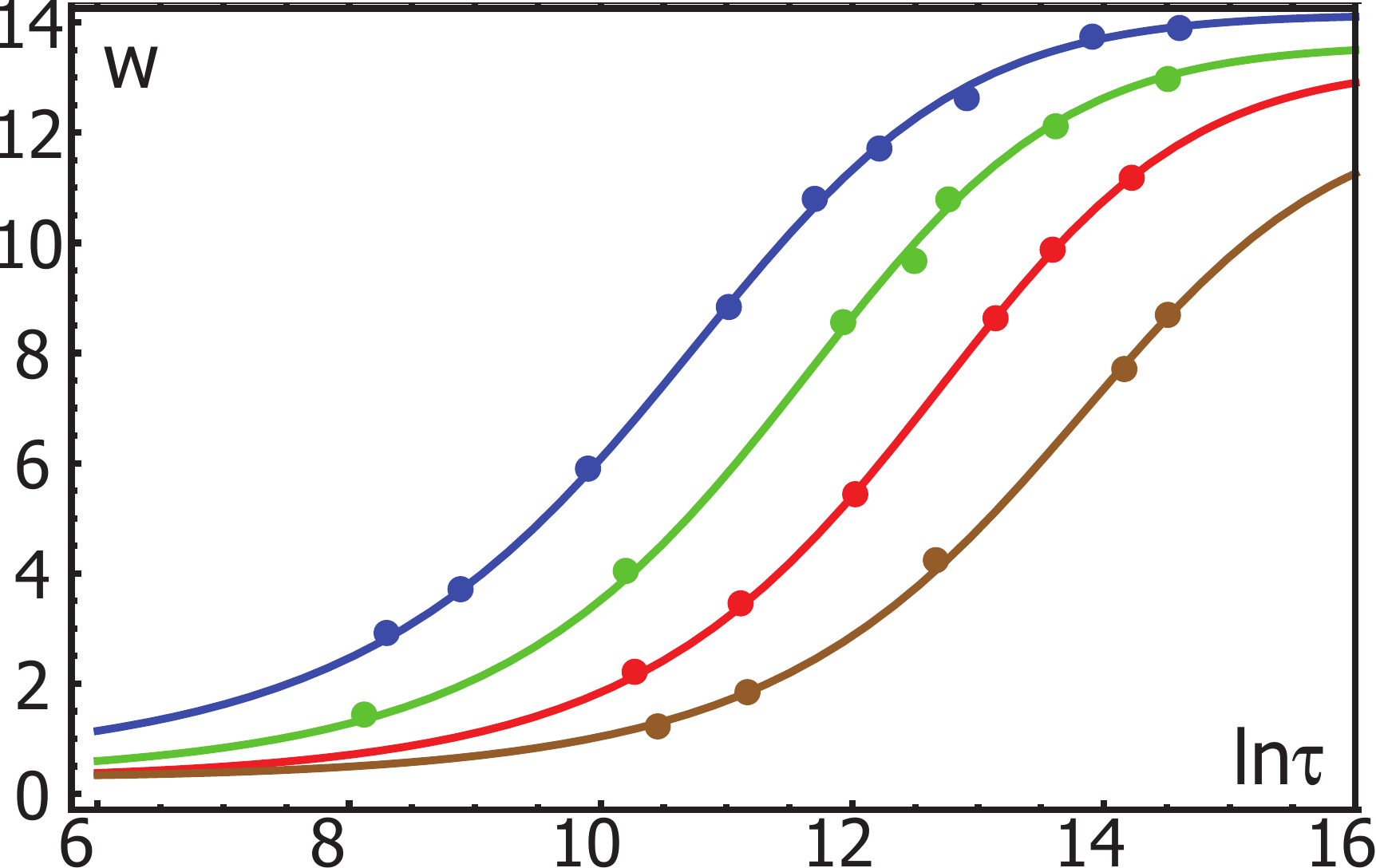}

\protect\caption{Time dependence of energy fluctuations for the subsystems of different
sizes $l=10,\,20,\,40,\,80$ (left to right) and their fits to log-normal
law (\ref{eq:w_tau}). The length of the full system was $L=5l$.
The best fit shown here corresponds to the value $\alpha=0.05$, $\beta/l=1.0-1.5$
and $\ln\tau_{0}\approx3.0-4.5$. The values for larger sizes ($\beta/l\approx1.0$
and $\ln\tau_{0}\approx4.5$ agree very well with the ones expected
for $R_{t}$ obtained in the computation of the resistance at $u=3.5:$$\ln R_{t}\approx5.0$.
This yields $r_{t}\approx5.0$ , $\beta/l\approx1.1$ and $\ln\tau_{0}\approx5.0$.
\label{fig:Time-dependence-of}}
\end{figure}

The numerical simulations confirm that the energy variance $w$ relaxes
in agreement with (\ref{eq:w_tau}) as shown in Fig. \ref{fig:Time-dependence-of}.
\textcolor{black}{The best fit to the} equation (\ref{eq:w_tau})
yields parameters\textcolor{red}{{} }close to the expected, $\ln R_{t}=\ln\rho(u)r_{t}$,
$r_{t}\approx u$. Extrapolation gives $w_{\infty}(\infty)\approx10.0$
of $w_{\infty}(l)$ to large $l$, which is significantly smaller
than thermodynamic value $w_{Th}=14.0$ at $u=3.5$. 

\begin{figure}[t!]
\includegraphics[width=0.9\columnwidth]{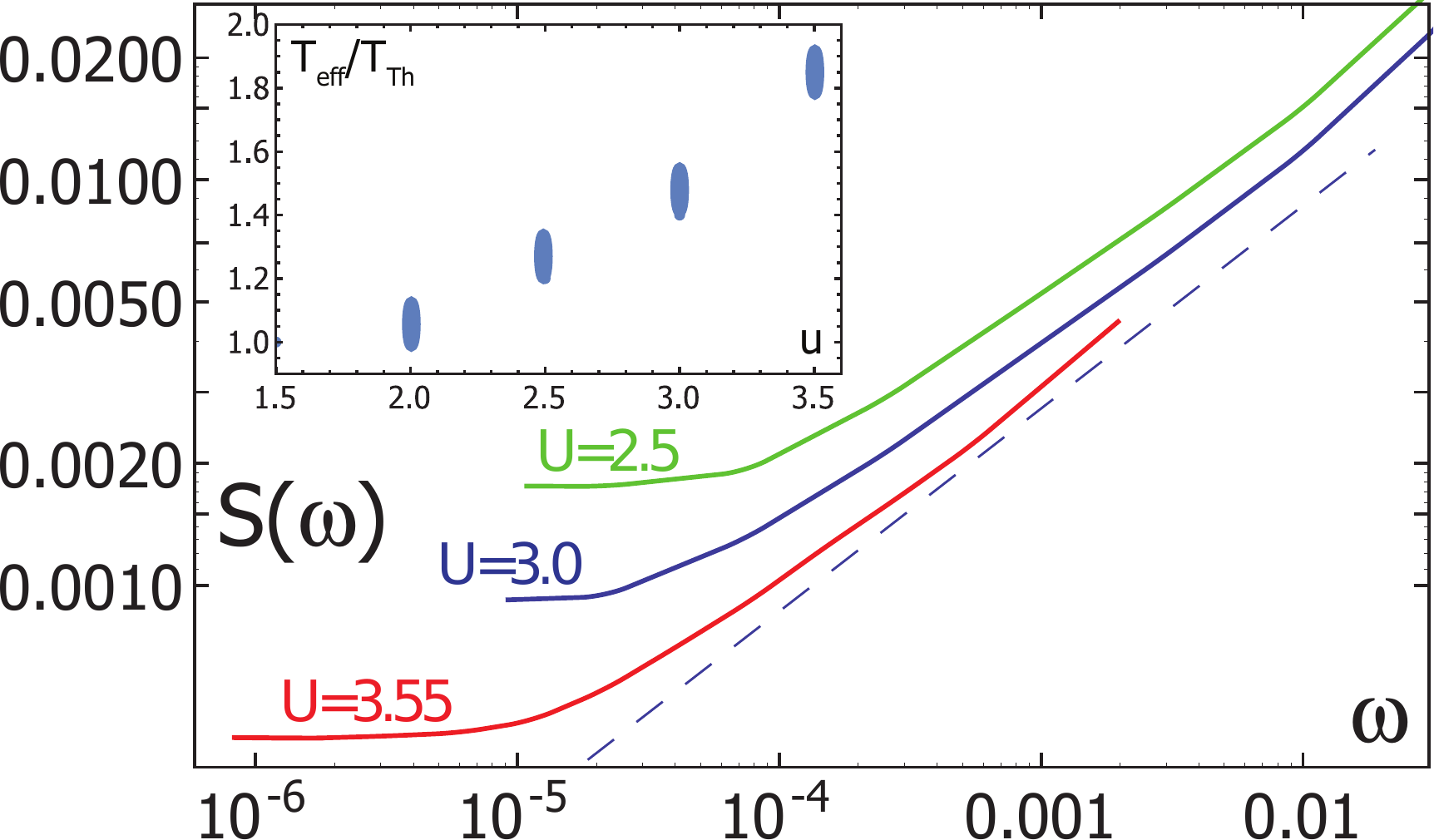}

\protect\caption{Spectrum of current noise fluctuations, $S(\omega)$ for different
internal energies, $U=2.5,\,3.0,\,3.55$. The dashed line shows $S(\omega)\sim\omega^{1/2}$
dependence. Insert shows the effective temperature determined from
FDT relation.\label{fig:Spectrum-of-current}}
\end{figure}

\emph{Another test of the ergodicity} follows from the fluctuation-dissipation
theorem (FDT) that relates conductivity and current fluctuations. In
the low frequency limit the noise power spectrum is $S(\omega\rightarrow0)=2T_{eff}/R$
where $T_{eff}$ is the effective temperature (see Fig.~\ref{fig:Spectrum-of-current}),
which we extracted from the numerical data. We found that $T_{eff}>T_{Th}$
for $u\gtrsim1$, where $T_{Th}$ is thermodynamic temperature. In
particular, $T_{eff}\approx1.6T_{Th}$ for $u=3.5$, which is close
to $T_{*}(u)$ shown on the inset to Fig. \ref{fig:Energy-fluctuations-as}. 

\emph{Quantum behavior.} In contrast to a classical limit $E_{C}\rightarrow0$
in the quantum regime $E_{C}>0$ we expect at $T=T_{c}$ a MBL phase
transition between two non-ergodic states: the insulator ($\rho=\infty$)
and a bad metal ($\rho<\infty$). For $E_{J}\gg E_{C}$ the bad metal
can be described classically at $T\ll T_{c}\sim E_{J}^{2}/E_{C}$.
Our previous discussion suggests that the bad metal is non-ergodic
in a broad range of the parameters, $T/E_{J}$ and $E_{J}/E_{C}$.
To verify this conjecture numerically we reduced the Hilbert space
of the model (\ref{eq:H}) to a finite number of charging states at
each site, $q_{i}=0,\pm1,\pm2$ (RHS model). We analyzed the time
evolution of entanglement entropy, $S\{\Psi\}$ of the left half of
the system. The entropy was averaged over the initial states from
the ensemble of product states in the charge basis, $S_{t}(L)\equiv<S\{\Psi\}>_{\Psi_{in}}$
that correspond to zero total charge. As a result, we obtained the
Gibbs entropy at $T=\infty$ (all states have the same weight $\exp(-H/T)\equiv1$)

\begin{figure}[t!]
\includegraphics[width=0.9\columnwidth]{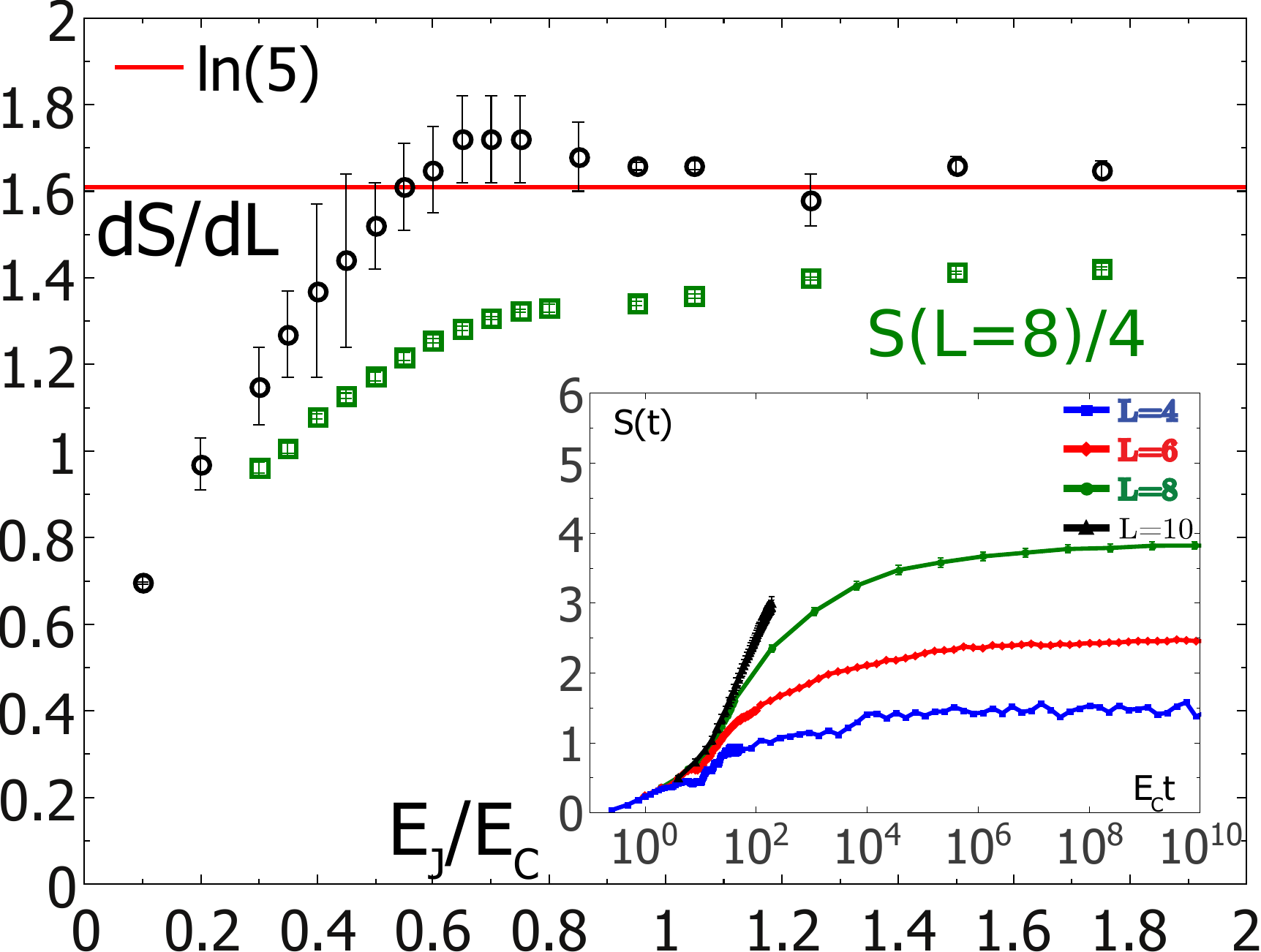}

\protect\caption{Enropy per spin in thermodynamic limit extrapolated to $t=\infty$.
The insert shows $S(t)$ dependence in the bad metal regime at $E_{J}/E_{C}=0.3$
that shows slow relaxation of the entropy to its saturation value.
\label{fig:Extrapolated-value-of} }

\end{figure}

The insert to Fig. \ref{fig:Extrapolated-value-of} shows the time
dependence of the entropy at $E_{J}/E_{C}=0.3$ which corresponds
to the bad metal regime (see below). A slow saturation of the entropy
follows its quick initial increase. It is crucial that the saturation
constant, $S_{\infty}(L)$ is significantly less than its maximal
value, $S_{Th}(L)=L\ln5$ expected at $T=\infty$ equilibrium. Furthermore,\textcolor{red}{{}
}$dS_{\infty}(L)/dL<\ln5$, indicating that $S_{\infty}-S_{Th}$ is
extensive and the system is essentially non-ergodic. 

Fig.~\ref{fig:Extrapolated-value-of} presents $S_{\infty}$ as a
function of $E_{J}/E_{C}$. Note that $S_{\infty}$ is measurably
less than $S_{Th}$ for $E_{J}/E_{C}<0.6-1$. For $E_{J}/E_{C}\gg1$,
the entropy saturation is quick and the accuracy of the simulations
does not allow us to distinguish $S_{\infty}$ from $S_{Th}$ (see
Supplementing material). We thus are unable to conclude if the system
is truly ergodic or weakly non-ergodic at $E_{J}/E_{C}\gg1$. The
former behavior would imply a genuine phase transition between bad
and good metals, while the latter corresponds to a crossover. 

Deep in the insulator the time dependence of the entropy is extremely
slow, roughly linear in $\ln t$ in a wide time interval (see Supplemental
materials). This resembles the results of the works \cite{Bardarson2012,Serbyn2013}\textbf{
}for the conventional disordered insulators. The extremely long relaxation
times can be attributed to rare pairs of almost degenerate states
localized within different halves of the system. The exponential decay
with distance of the tunneling amplitude that entangles them leads
to the exponentially slow relaxation. 

\begin{figure}[t!]
\includegraphics[width=0.9\columnwidth]{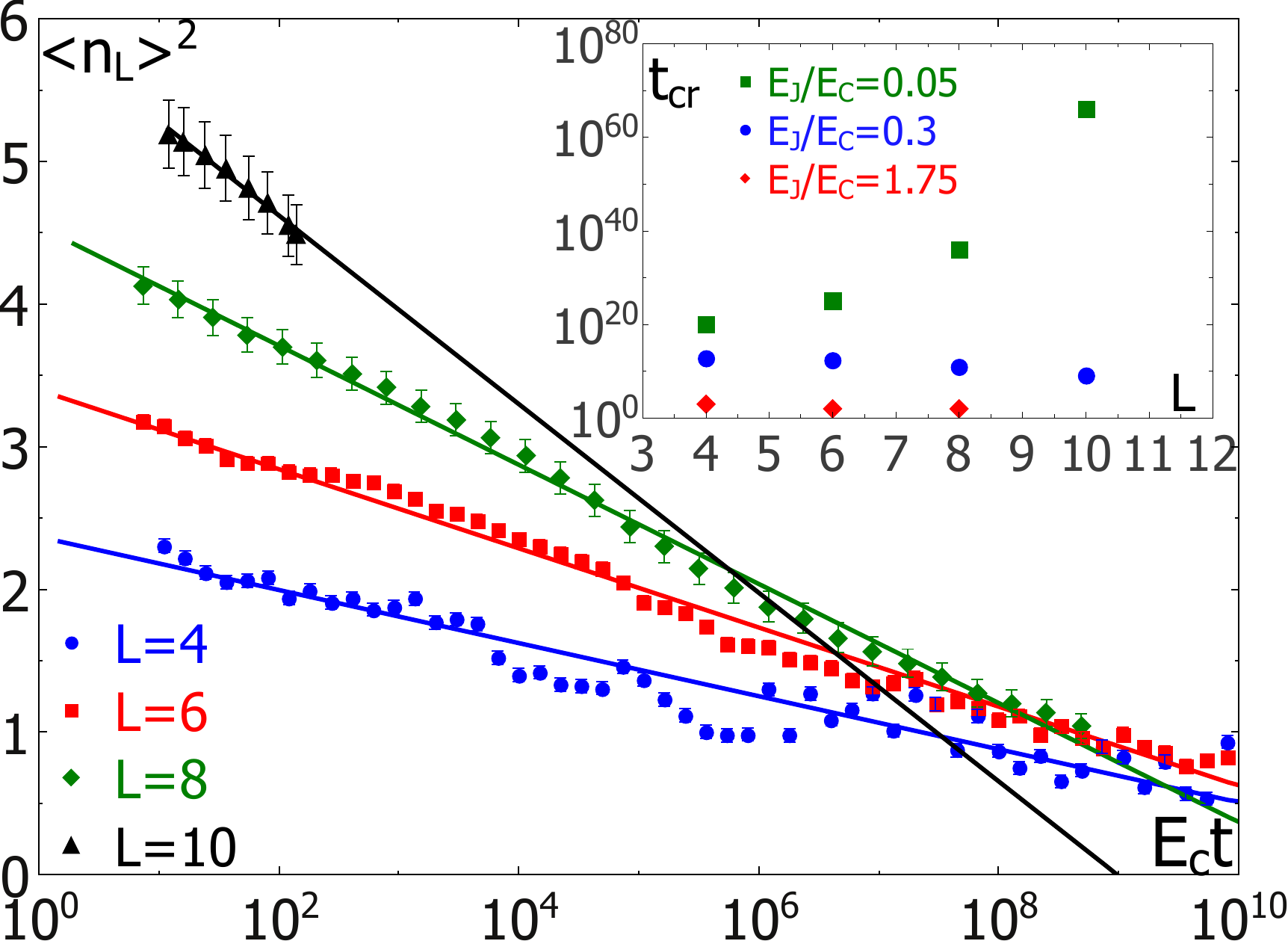}

\protect\caption{Charge relaxation in a bad metal ($E_{J}/E_{C}=0.3)$. The characteristic
time scale, $t_{cr}$, can be defined by the extrapolated crossing
point of $\left\langle n^{2}\right\rangle $ with t-axis. The crossing
point $t_{cr}$ shows dramatically different behavior for the insulator
and metal as shown by insert: in the good metal this time is very
short, in the bad metal it is dramatically longer but does not increase
with size whilst in insulator it is extremely long and grows with
size. \label{fig:Charge-relaxation-in-bad-metal}}
\end{figure}

In order to locate the MBL transition we analyzed the time dependence
of the charge fluctuations. In a metal the charge fluctuations relaxation
rate depends weakly (as a power law) on the sample size in contrast
to the exponential dependence in the insulator. Comparing the dependences
of the rates on the system size for different $E_{J}/E_{C}$ (see
Fig. \ref{fig:Charge-relaxation-in-bad-metal} ) we see that the transition
happens in the interval $0.05<E_{J}/E_{C}<0.3$. 

The variances of the charge in the RHS model at $T=\infty$ and the
problem (\ref{eq:H}) at finite $T$ coincide at $T=2E_{C}$. Thus,
we expect that the results of the quantum simulations describe the
behavior of the model (\ref{eq:H}) at $T/E_{J}\sim E_{C}/E_{J}$
yielding the hyperbola shown in Fig. \ref{fig:Phase-diagram-of}.
The MBL transition at $E_{J}/E_{C}\sim0.2$ in $T=\infty$ RHS model
corresponds to the transition temperature $T_{c}\sim10E_{J}$ in model
(\ref{eq:H}). The transition line shown in Fig. \ref{fig:Phase-diagram-of}
is a natural connection of this point with $T_{c}\approx E_{J}^{2}/E_{C}$
asymptotic at $E_{J}/E_{C}\gg1$ discussed above. The maximum of the
transition temperature is natural because the quantum fluctuations
are largest at $E_{J}\sim E_{C}.$

\emph{Possible experimental realization.} MBL and the violation of
the ergodicity can be observed only at sufficiently low temperatures
when one can neglect the effects of thermally excited quasiparticles
which form the environment to model (\ref{eq:H}). This limits temperatures
to $T\lesssim0.1\Delta$ where $\Delta$ is superconducting gap. In
order to explore the phase diagram one has to vary both $T/E_{J}$
and $E_{J}/E_{C}$ in the intervals $1\lesssim T/E_{J}\lesssim5$
and $0.1\lesssim E_{J}/E_{C}\lesssim5$. The former condition can
be satisfied if each junction is implemented as a SQUID loop with
individual Josephson energy $E_{J}^{(0)}\sim T_{max}=0.1\Delta$ so
that $E_{J}=2\cos(e\Phi/\pi\hbar)E_{J}$ where $\Phi$ is flux through
the loop. The latter condition requires enhancing ground capacitance
of each island which should exceed the capacitance of the junctions
in order for the model (\ref{eq:H}) to be relevant. Realistic measurements
of such array include resistance $R(T,E_{J})$ and current noise.
Here we predict a fast growth with temperature and divergence at $T_{c}$
of the resistivity and violation of FDT. 

In conclusion, we presented strong numerical evidences for the MBL
transition and its semi-quantitative description in a \emph{regular},
disorder free%
\footnote{When this work was finished we learnt about the numerical studies\cite{Moore2014,Silva2014}
that reports MBL in disorder free 1D systems, in cotrast the work
\cite{Abanin2015} reports the absence of MBL in such systems. All
these works studied models different from ours. %
} Josephson chain. Probably the most exciting finding is the intermediate
non-ergodic conducting phase (bad metal) between the MBL insulator
and good ergodic metal. This phase distinguishes Josephson junction
chain from the spin$1/2$ Heisenberg one-dimensional model in the
random field, \cite{Oganesyan2007,Oganesyan2009,Luitz2014}, where
the ergodicity is believed to be violated \emph{only }in the MBL regime\cite{Luitz2014}.
The ergodicity violation in a wide range of parameters is in contrast
to the one particle Anderson localization in a finite-dimensional
space where the non-ergodic behavior is limited to the critical point.
However, the signatures of the ``non-Gibbs regime'' in the discrete
nonlinear Schrodinger equation \cite{Rasmussen2000,Flach2008}, the
appearance of the non-ergodic states in a single particle problem
on the Bethe lattice as well as the dominance of single path relaxation
close to the critical point of superconductor-insulator transitions
on Bethe lattice\cite{Feigelman2010,Cuevas2012} make us believe that
the intermediate non-ergodic phase is a generic property of MBL transition
rather than an exception. Further work is needed to establish the
domain of the applicability of these results as well as the nature
of the transition between bad and good metals. 

\textbf{Methods}

\emph{Simulation of the JJA in the classical regime.} At large $u$
averaging out temporal fluctuations requires exponentially long times.
Moreover, at $u\gg1$ the resistance increases with $u$ factorially
leading to a strong heating in the computation of resistance unless
the measurement current is factorially small. Observation of a small
current against the background of a low frequency noise requires increasingly
long times. Accordingly, for the realistic evolution times $\tau_{ev}\lesssim10^{8}$
the resistance can be computed only for $u<4.0$. 

\emph{Simulation of the quantum problem. }The time dependence of the
entropy and the charge fluctuations for system of sizes $L=4,6,8$
was computed using exact diagonalization in a symmetric subspace under
charge conjugation. The tDMRG method was employed for larger sizes
which accuracy limits the range of times that we could study. In all
simulations we impose the particle number conservation and open boundary
conditions.

\bibliographystyle{unsrt}
\bibliography{MBLandNonequilibrium}

\textbf{Acknowledgment}

We are grateful to I.L. Aleiner, M. Feigelman, S. Flach, V.E. Kravtsov
and A.M. Polyakov for useful discussions. The work was supported in
part by grants from the Templeton Foundation (40381), ARO (W911NF-13-1-0431)
and ANR QuDec. 

\newpage{}

\part*{Supplemental materials}

\section{Free energy of the chain in thermodynamic equilibrium}

Evaluation of the free energy with Hamiltonian (\ref{eq:H}) gives
\[
F=\frac{1}{2}T\ln\left(\frac{E_{C}}{2\pi T}\right)-T\ln\left[I_{0}(E_{J}/T)\right]
\]
which allows us to express $u$ in terms of $T$: 
\begin{equation}
u=\frac{T}{2E_{J}}+\left(1-\frac{I_{1}(E_{J}/T)}{I_{0}(E_{J}/T)}\right)\label{eq:u_Th}
\end{equation}
 Here $I_{\alpha}(x)$ are modified Bessel functions. Using (\ref{eq:u_Th})
one can check the validity of (\ref{eq:U^2}), which takes the form
$w=T^{2}du/dT$ .

\section{Noise distribution. }

At high temperatures the frequencies of individual phases are typically
large, $\omega_{i}\gg1$ which allows to solve the equations (\ref{eq:Eq_of_motion})
in perturbation theory. Furthermore, because the effect of noise decreases
exponentially with distance one can use the forward propagation approximation
in which the evolution of the next phase is determined exclusively
by the previous one
\begin{equation}
\frac{d^{2}\phi_{i}}{d\tau^{2}}=-\sin(\phi_{i}-\phi_{i-1})\label{eq:ClForwardProp}
\end{equation}

Looking for the solution in the form $\phi_{i}=\omega_{i}\tau+\delta\phi_{i}(\tau)$
and introducing the notation 
\[
e^{i\phi_{i}}=\int d\omega e^{-i\omega\tau}z_{i}(\omega)
\]
for the noise at site $i$ we get 
\begin{equation}
\delta\phi_{i}=\frac{1}{(\omega_{i}-\omega)^{2}}\mbox{Im}z_{i-1}(\omega)e^{i(\omega-\omega_{i})\tau}\label{eq:deltaphi}
\end{equation}
We are interested in the propagation of the low frequency harmonics
of the noise,$\omega\ll\omega_{i}$. Neglecting the high frequency
components of the noise one obtains from (\ref{eq:deltaphi}) the
noise recursion 
\begin{equation}
z_{i}(\omega)=\frac{1}{2(\omega_{i}-\omega)^{2}}z_{i-1}(\omega)\label{eq:NoiseRecursion}
\end{equation}
Iterations of (\ref{eq:NoiseRecursion}) in the limit $\omega\ll\omega_{i}$
lead to the equation (5) of the main text. The exponential decrease
of the noise away from its source in the classical regime is in agreement
with the numerical computation in the quantum regime (see section
\ref{sub:Charge-propagation-in}). 

The recursion (5) of the main text implies that the noise generated
by a single triad at distance, $r\gg1$ is given by a product of a
large number of factors. Being a sum of a large number of random independent
terms the logarithm of the noise has Gaussian distribution. Thus,
at the distance $r_{t}\gg r\gg1$, from the closest triad the noise
distribution has a log normal form
\begin{equation}
P_{r}(\zeta)=\frac{1}{\sqrt{2\pi\zeta_{*}(r)}}\exp\left[-\frac{\left(\zeta-\zeta_{*}(r)\right)^{2}}{2\zeta_{*}(r)}\right]\label{eq:P(lnz)-1}
\end{equation}
where $\zeta=\ln z$, $\zeta_{*}(r)=\left(\gamma+\ln u\right)r$,
and $\gamma=0.577$ is Euler constant. The exponential decay of the
effect of a single triad (\ref{eq:z_i+k},\ref{eq:P(lnz)-1}) implies
the charge localization. The DC charge transport in a macroscopically
large array requires interaction of different triads through the quiet
regions. A quiet interval $(j,j+r)$ between islands can be characterized
by the resistance $R_{j,j+r}\sim z_{j}/z_{j+r}$. Convolution of distribution
(\ref{eq:P(lnz)-1}) with the Poisson distribution for the sizes of
the quiet regions, $r$, yields the log-normal distribution for the
resistances $R_{j,j+r}$ of quiet regions.

\section{Charge propagation in quantum insulating phase. \label{sub:Charge-propagation-in}}

\begin{figure}
\includegraphics[width=0.9\columnwidth]{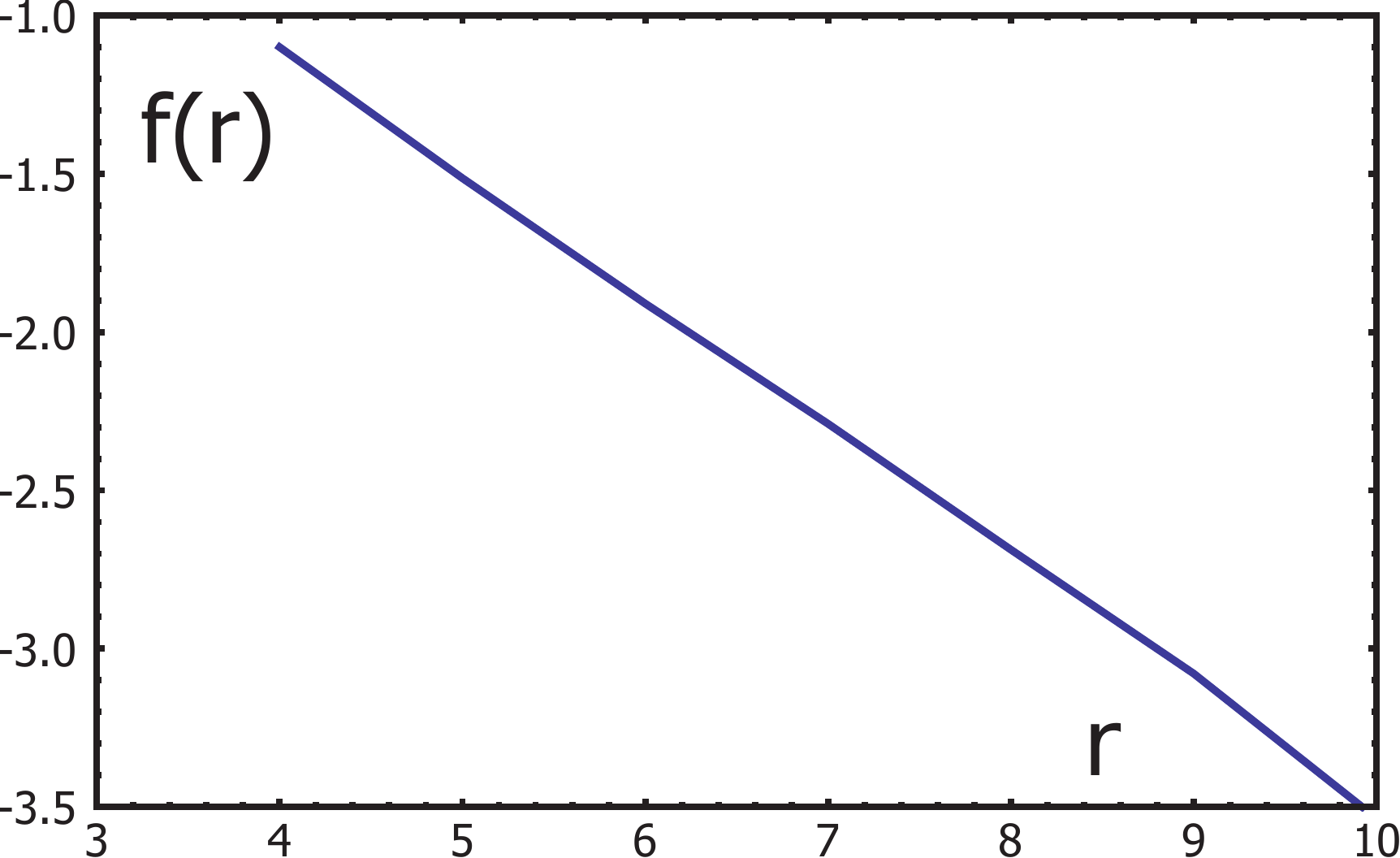}

\protect\caption{Logarithm of the dimensionless amplitude of the charge transfer, $f(r)$
determined in the forward propagation approximation.\label{fig:Logarithm-of-the}}
\end{figure}

In the insulating phase the charge transfer by distance $r$ appears
in the $r^{th}$ order of the perturbation theory in the parameter
$E_{J}/\sqrt{TE_{c}}.$ For small $E_{J}/\sqrt{TE_{c}}\ll1$ the main
contribution to the amplitude, $\Psi(r)$ of this processes comes
from exactly $r-1$ virtual hops between neighboring sites in the
same direction (forward propagation approximation). These hopes can
occur in any order. Each sequence of the hops corresponds to a particular
permutation, $P$, of $1\ldots r-1$. After $n<r-1$ hops the charging
energy changes by $\Delta E_{n}=E_{c}\sum_{k\leq n}(q_{P(k)+1}-q_{P(k)})$,
so that 
\[
\left\langle \ln\left|\Psi_{FWD}(r)\right|\right\rangle =\left\langle \ln\sum_{P}\frac{E_{J}}{2E_{c}\left|\sum_{k}(q_{P(k)+1}-q_{P(k)})\right|}\right\rangle _{q}
\]
where $\left\langle \ldots\right\rangle _{q}$ means average over
charge configurations normally distributed with variance $T/E_{C}.$
Rescaling the charge and factoring out the dimensionfull parameters
we find that 
\begin{equation}
\left\langle \ln\left|\Psi_{FWD}(r)\right|\right\rangle =r\ln\left(\frac{E_{J}}{\sqrt{TE_{c}}}\right)+f(r)\label{eq:lnPsi_FWD}
\end{equation}

\begin{equation}
f(r)=\left\langle \ln\sum_{P}\frac{1}{2\left|\sum_{k}(q_{P(k)+1}-q_{P(k)})\right|}\right\rangle _{q}\label{eq:f(r)}
\end{equation}
where average is performed over charge configurations with unit variance.
The numerical evaluation of $f(r)$ shows that $f(r)\approx\eta r$
with $\eta\approx-0.3$(see Fig. \ref{fig:Logarithm-of-the}). 

We estimate the transition temperature by demanding $\left\langle \ln\left|\Psi_{FWD}(r)\right|\right\rangle \sim const$
which gives $T_{c}^{FWD}\approx e^{2\eta}E_{J}^{2}/E_{C}$. The forward
propagation approximation neglects the effects of the backscattering
and level repulsion. All these apparently small effects favor the
insulating behavior, thus, the actual transition is below $T_{c}\lesssim T_{c}^{FWD}$.
We conclude that the numerical \textcolor{green}{pre}factor, $\tilde{\eta}$,
in the equation for the critical temperature, $T_{c}=\tilde{\eta}E_{J}^{2}/E_{C}$
is close to one, $\tilde{\eta}\lesssim0.5$.

The estimate of the charge transfer (\ref{eq:lnPsi_FWD}) neglects
the charge discreteness which is expected to become irrelevant at
$T/E_{C}\rightarrow\infty$. To estimate the effect of the discreteness
at finite $T/E_{C}$ we note that it is dominated by the appearance
of exactly zero $\Delta E_{n}$ for some $n$ (resonances). Such degeneracies
are lifted in the next order of the perturbation theory: the energy
difference between the states with charge $q_{i}+1$ and charge $q_{i}$
at island $i$ becomes
\begin{eqnarray*}
E(q_{i}+1)-E(q_{i}) & = & E_{C}(q_{i}+\frac{1}{2})+\delta_{i}\frac{E_{J}^{2}}{E_{C}}\\
\delta_{i} & = & \delta_{i,i+1}+\delta_{i-1,i}
\end{eqnarray*}

\[
\delta_{i,i+1}=\frac{q_{i+1}-q_{i}-\frac{1}{2}}{\left[\frac{3}{4}-\left(q_{i+1}-q_{i}-\frac{1}{2}\right)^{2}\right]^{2}-\left(q_{i+1}-q_{i}-\frac{1}{2}\right)^{2}}
\]

Close to the transition line we estimate $\delta_{i}\sim(E_{C}/E_{J})^{3}$,
so the degeneracy is lifted by $\delta E\sim E_{C}^{2}/E_{J}\ll E_{J}$.
These resonances occur with the probability $E_{C}/E_{J}$, at an
average distance $r\sim E_{J}/E_{C}$ from each other. Close to the
transition the hopping amplitude between these sites is given by 
\[
\Psi(r)\sim E_{J}\exp\left(-\frac{T-T_{c}}{2T_{c}}r\right)
\]
The energy difference, $\Delta E$, exceeds this amplitude provided
that 
\begin{equation}
\frac{T-T_{c}}{T_{c}}>\gamma'\frac{E_{C}}{E_{J}}\ln\frac{E_{J}}{E_{C}}\label{eq:T-T_c}
\end{equation}
 where $\gamma\sim1$ is numerical coefficient. The many body localization
takes place only when the condition (\ref{eq:T-T_c}) is satisfied.
We conclude that the shift of the transition line upwards is small
in $E_{C}/E_{J}$ (\ref{eq:T-T_c}). 

According to (\ref{eq:f(r)}) the charge propagation is controlled
by the average logarithm of the charge difference. This suggests that
the simulations in which for the charge is evenly distributed in the
interval $(-Q,Q)$ should resemble the simulations for the normal
distribution with variance $T/E_{C}$ when the two cases correspond
to the same$\left\langle \ln q_{i}\right\rangle $ or $\left\langle \ln(q_{i}-q_{i+1})\right\rangle .$
Application of such criteria leads to the conclusion that restriction
of the charges$-Q\leq q_{i}\leq Q)$ at $T\to\infty$t is equivalent
to the unrestricted charges (\ref{eq:H}) with effective temperature
$T=\gamma'E_{C}Q^{2}$ with $\gamma'\approx1$ .

\section{Charge and entropy relaxation in quantum regime}

\begin{figure}
\includegraphics[width=0.9\columnwidth]{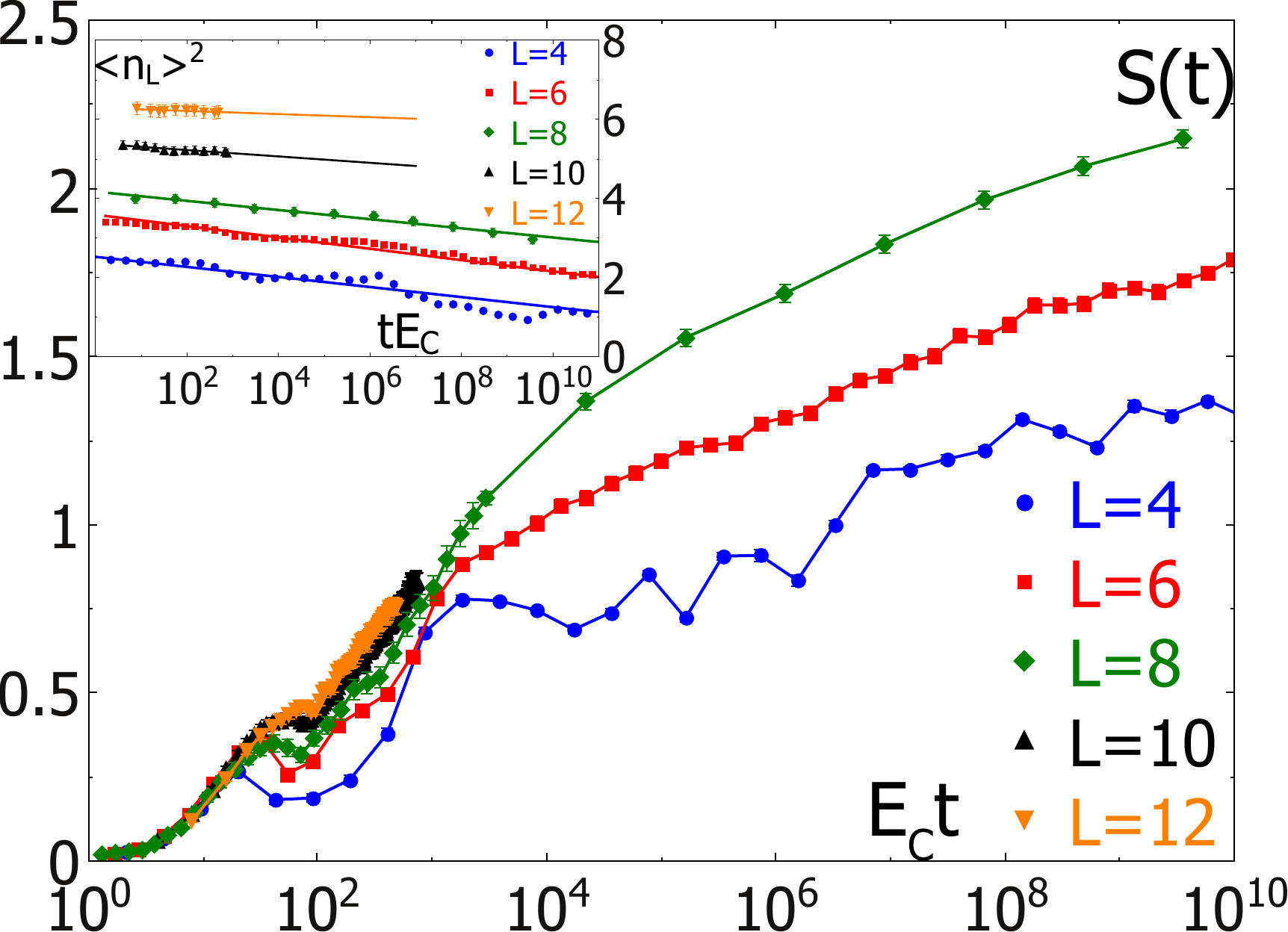}

\protect\caption{Time evolution in the insulator, $E_{J}/E_{C}=0.05$. The main panel
shows slow evolution of the entropy which is roughly linear in $\ln t$
in a wide range of times. The insert shows slow relaxation of the
charge which becomes even slower for longer samples. \label{fig:Time-dependence-of-entropy-insulator}}
\end{figure}

Here we give the details of the numerical results for charge and entropy
evolution in the insulating and metallic regimes that are both qualitatively
different from the non-ergodic bad metal phase. 

We begin with the \emph{insulating phase}. The behavior of the entropy
in a wide time range ($t<10^{10}$) is shown in Fig. \ref{fig:Time-dependence-of-entropy-insulator};
one can distinguish three distinct regimes. 

At shortest times ($E_{C}t\lesssim20)$ the entropy growth fast. The
fact that this growth is identical for the systems of different sizes
suggests that it is due to the particles spreading across the boundary
which happens at time scales $t\sim1/E_{J}$. In the intermediate
time region ($20\lesssim E_{C}t\lesssim10^{3}$) the entropy is almost
size independent for all but smallest sizes and small ($S\lesssim1$).
Similar behavior is observed for disordered systems; it is due to
a small entanglement of typical excitations a bit further from the
boundary or to a larger entanglement of rare excitations with exceptionally
close energies. Because the amplitude of the entanglement decreases
exponentially fast with distance the contribution to the entropy coming
from the states far from the boundary is small and thus the entropy
demonstrates weak size dependence in this regime. 

The striking feature of the insulating state in this model is a very
slow (logarithmic) growth of entropy at long times ($E_{C}t>10^{4}$)
that remains much below $\ln5$ per site. We attribute this dynamics
to degenerate charge configurations: even when located far from each
other such configurations can hybridize due to a small but non zero
$E_{J}/E_{C}$. For example, configurations connected by inversion
symmetry possess a matrix element which is exponentially small in
system size. The time needed to resolve this couple is thus exponentially
large. 

\begin{figure}
\includegraphics[width=0.9\columnwidth]{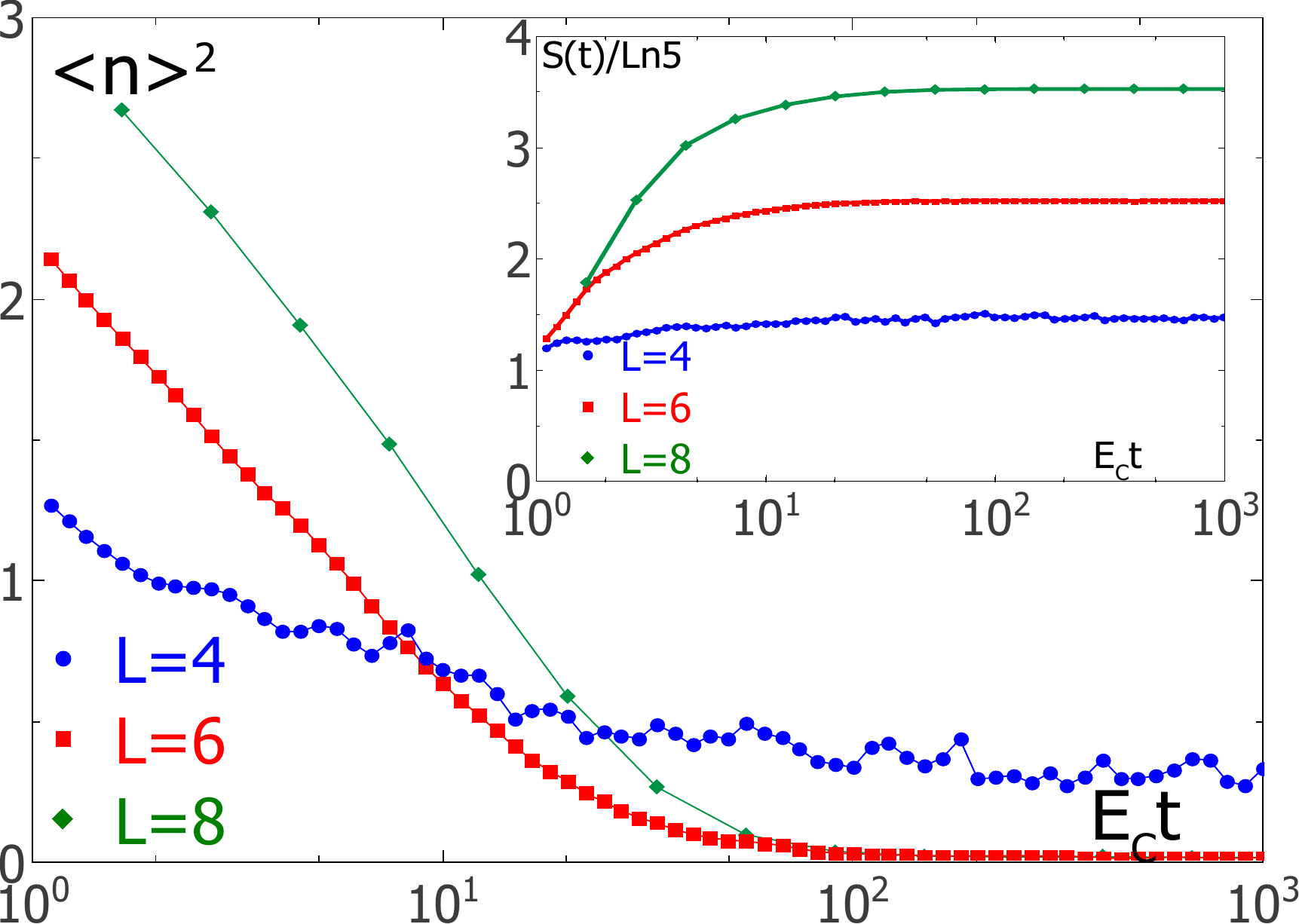}

\protect\caption{Charge relaxation and entanglement entropy relaxation in a good metal
realized at $E_{J}/E_{C}=1.75$\label{fig:Entanlement-entropy-in-good-metal}}
\end{figure}

In contrast to the entropy, the charge fluctuations display simple
monotonic behavior at all times. The characteristic charge relaxation
time increases extremely rapidly with the system size. We conclude
that this phase is a genuine insulator which phase space is separated
into thermodynamically large number of independent compartments. 

\emph{In a good metal} there is only one regime for charge and entropy
time evolution. The charge relaxes quickly and this relaxation does
not show any sign of getting slower at larger system sizes, see Fig.
\ref{fig:Entanlement-entropy-in-good-metal}. Accordingly, the entropy
increases rapidly and saturates at the values that approach $L\ln5$
for large system sizes.

\section{Berezinski-Kosterlitz-Thouless critical point at zero temperature}

In this section we give the details of the numerical methods that
allowed us to determine the value of the ratio of Josephson to charging
energies, $\eta$, for the Berezinsky-Kosterlitz-Thouless phase transition
of the model (\ref{eq:H}) at zero temperature. Finite order derivatives
of the energy do not display any discontinuity at this transition.
Thus, we have to use more sophisticated numerical methods than the
ones used in the case of second order phase transitions.

\begin{figure}[!t]
\includegraphics[width=0.9\columnwidth]{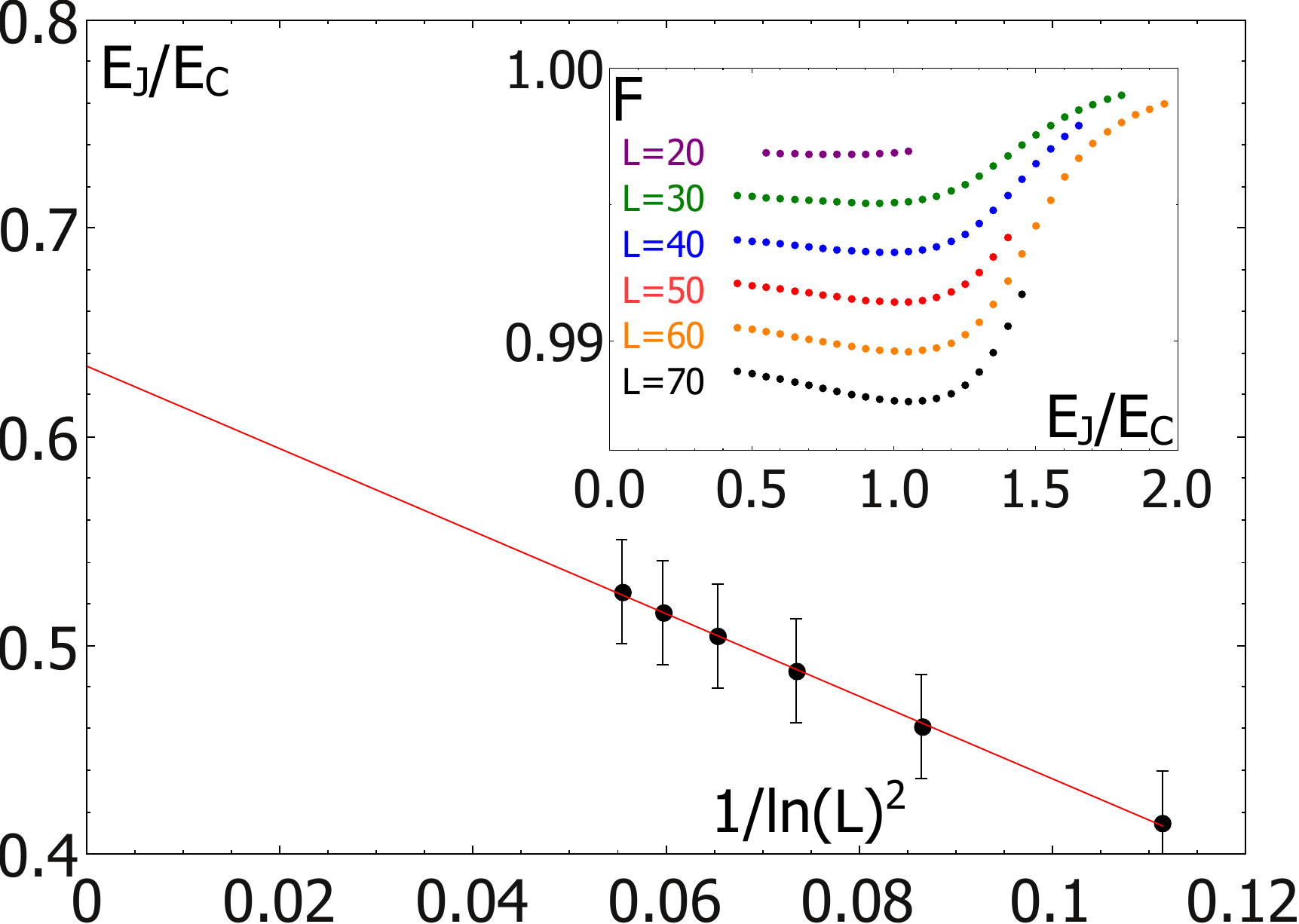}

\protect\caption{The insert shows the fidelity of the ground-state as a function of
$E_{J}/E_{C}$ for different sizes $L$. The minimum of the fidelity
as a function of $[ln(L)]^{-2}$ appear in the main panel. The red
line fits the data to a line. The location of the critical point can
be extracted from the value of the fitting at origin $E_{J}/E_{C}=0.63\pm0.04$.\label{fig:BKT}}
\end{figure}

The idea of our approach is to compute the fidelity of the ground-state
defined by the equation 
\begin{equation}
F(E_{J}/E_{c})=<\psi_{gs}(E_{J}/E_{c}+\delta)|\psi_{gs}(E_{J}/E_{c})>\label{eq:Fidelity}
\end{equation}
for small $\delta$.
The result obtained by DMRG method \cite{White92} is shown in the
insert of Fig. \ref{fig:BKT} as a function of $E_{J}/E_{c}$ for
different sizes. One expects that fidelity is minimal at the critical
point \cite{Buonsante07}. In the main panel of Fig. \ref{fig:BKT}
we show positions of this minima as well as its extrapolation to infinite
size using $E_{J}/E_{C}(L)=a[ln(L)]^{-2}+\eta$ \cite{Pino2012}.
This procedure yields $\eta=0.63\pm0.04$ cited in the main text. 
\end{document}